\documentclass[runningheads]{llncs}
\usepackage[T1]{fontenc}
\usepackage{multirow}
\usepackage{makecell}
\usepackage{xcolor}
\usepackage{amssymb}
\usepackage{graphicx,verbatim}
\begin{document}
\title{MM2CT: MR-to-CT translation for multi-modal image fusion with mamba}
%

\author{Chaohui Gong\textsuperscript{1},
\ Zhiying Wu\textsuperscript{1*}, 
Zisheng Huang\textsuperscript{1}, 
Gaofeng Meng\textsuperscript{1},\\ 
Zhen Lei\textsuperscript{1}, 
and Hongbin Liu\textsuperscript{1}}  
\authorrunning{C. Gong et al.}
\institute{
    \textsuperscript{1}Centre for Artificial Intelligence and Robotics, Hong Kong Institute of Science \& Innovation, Chinese Academy of Sciences \\
    \email{zhiying.wu@cair-cas.org.hk} \\
}

    
\maketitle              
\begin{abstract}
Magnetic resonance (MR)-to-computed tomography (CT) translation offers significant advantages, including the elimination of radiation exposure associated with CT scans and the mitigation of imaging artifacts caused by patient motion. The existing approaches are based on single-modality MR-to-CT translation, with limited research exploring multimodal fusion. To address this limitation, we introduce Multi-modal MR to CT (MM2CT) translation method by leveraging multimodal T1- and T2-weighted MRI data, an innovative Mamba-based framework for multi-modal medical image synthesis. Mamba effectively overcomes the limited local receptive field in CNNs and the high computational complexity issues in Transformers. MM2CT leverages this advantage to maintain long-range dependencies modeling capabilities while achieving multi-modal MR feature integration. Additionally, we incorporate a dynamic local convolution module and a dynamic enhancement module to improve MRI-to-CT synthesis. The experiments on a public pelvis dataset demonstrate that MM2CT achieves state-of-the-art performance in terms of Structural Similarity Index Measure (SSIM) and Peak Signal-to-Noise Ratio (PSNR). Our code is publicly available at https://github.com/Gots-ch/MM2CT. 


\end{abstract}
\section{Introduction}
In clinical practice, medical image such as Magnetic Resonance Imaging (MRI), Computed Tomography (CT) are crucial as it provides information about the human body~\cite{dayarathna2023deep}. The radiation exposure during CT scanning~\cite{bosch2023risk} has prompted researchers to explore image synthesis technology as an alternative solution. Medical image synthesis technology aims to predict imaging manifestations of one modality using data from another modality. However, different imaging devices operate on distinct physical principles~\cite{armanious2020medgan,DBLP:journals/bspc/WuLZWWLLL24,F2PASeg2025miccai}. This leads to significant nonlinear differences in tissue contrast. Consequently, cross-modal synthesis methods still facer challenges in feature space mapping. Currently, research is largely limited to image synthesis within a single modality. However, complementary features carried by different modality images can provide significant synergistic enhancement for anatomical structure analysis and pathological feature recognition~\cite{zhang2023multi,DBLP:conf/miccai/XuWCCLL24,DBLP:conf/iros/ChenZGL0W0L24}. MRI datasets usually consist of images from multiple imaging protocols~\cite{li2019diamondgan}, which provides conditions for multi-modal medical image fusion.

Image fusion is a specific algorithm that combines two or more images into a new image~\cite{li2021medical,FT-Reg2025}. The current mainstream methods mainly rely on basic operations like feature superposition and channel concatenation for multi-modal information integration. For example, WavTrans~\cite{li2022wavtrans} and MMHCA~\cite{georgescu2023multimodal} both concatenate the input tensors in the medical image super-resolution task. However, these operations have obvious limitations. First, WavTrans~\cite{li2022wavtrans} and MMHCA~\cite{georgescu2023multimodal} struggle to deeply explore and fully utilize the inherent correlational features between cross-modal data. Second, even with the introduction of attention mechanisms or convolutional neural networks, technical bottlenecks remain. These include inefficient inter-modal interactions and insufficient modeling of long-range dependencies.

To address these challenges, we propose MM2CT (Multi-Modal MR-to-CT), a novel deep learning framework based on state space modeling for MRI-to-CT cross-modal synthesis. The framework innovatively integrates Mamba for multimodal fusion. First, shallow features are extracted using convolutional layers and Mamba blocks. Subsequently, we employ channel swapping and a Mamba fusion module to effectively extract local detail features from different modalities. Following this processing, the features undergo further refinement through Mamba blocks and convolutional layers to generate the final synthesized image. The main contributions are summarized as follows:
\begin{enumerate}
 \item We first introduce the MM2CT model, a CT translation framework capable of simultaneously processing multimodal MR images. This model integrates complementary features from T1- and T2-weighted modalities to achieve high-quality unsupervised conversion between MR and CT images.
    \item A Mamba-based fusion module is designed to effectively capture image features. Combined with a dynamic enhancement module, the proposed MM2CT efficiently handles inter-modality differences to enhance texture information perception. 
    \item Our method outperforms existing state-of-the-art approaches across all evaluation metrics in the pelvic dataset. 
\end{enumerate}

\section{Related Work}
\subsection{Medical Image Translation}
Image-to-image translation technology based on deep learning provides a new paradigm for medical imaging modality conversion~\cite{MRtoCT2025isbi}. Early research mainly focused on the generative adversarial network (GAN) framework. These methods capture distribution features of target modalities through adversarial learning. This significantly improves the preservation of structural details [1]. For example, CycleGAN uses cycle consistency loss to deal with the problem of unpaired data and has been successfully applied to MRI-to-CT conversion [2]. MaskGAN~\cite{phan2023structure} innovatively introduces automatically extracted anatomical structure masks as prior knowledge. This enhances the anatomical structure accuracy and consistency of generated images. However, GAN-based methods generally suffer from unstable adversarial training~\cite{li2022new}. This may lead to artifacts or local detail distortion in generated images. 

In recent years, diffusion models have made significant breakthroughs in the field of medical image generation. Notably, Syndiff~\cite{ozbey2023unsupervised} pioneered an innovative approach by combining adversarial training with the diffusion framework, demonstrating exceptional performance in MRI-to-CT translation tasks. However, existing research has primarily focused on single-modality translation, which somewhat limits the full utilization of medical imaging information. Image fusion techniques thus present a promising solution to address this challenge.

\subsection{Multi-modal Image Translation}
Multi-modal image fusion aims to integrate complementary information from different modalities to generate high-quality fused images with rich textural details~\cite{dalmaz2022resvit,atli2024i2i}. With the advancement of deep learning technologies, neural network-based multi-modal fusion methods have achieved efficient cross-modal information integration through their powerful nonlinear modeling capabilities. For instance, Wu et al.~\cite{wu2024towards} proposed an innovative framework combining convolutional operations and attention mechanisms, which was successfully applied to prostate cancer classification using multi-modal transrectal ultrasound images.

Although Transformer models excel at capturing long-range dependencies, their practical application is limited by computational complexity that grows quadratically with sequence length~\cite{qu2024survey}. Recent studies have shown that architectures based on State Space Models (SSM) maintain model performance while significantly improving computational efficiency due to their lower computational complexity. In particular, the Visual State Model (Vmamba) ~\cite{liu2025vmamba} has been successfully applied to computer vision tasks, opening new technical pathways for medical image fusion.

Building on this foundation, the Mamba-based dual-phase fusion model (Mambadfuse)~\cite{li2024mambadfuse} innovatively proposes a dual-phase feature fusion module that effectively captures and integrates complementary information between different modalities, advancing multi-modal medical image fusion technology. Additionally, FusionMamba~\cite{xie2024fusionmamba} enhances multi-modal image fusion performance through a novel dynamic feature enhancement method, demonstrating significant advantages in biomedical image processing.

\begin{figure*}[t]
\centering
\includegraphics[width=\textwidth]{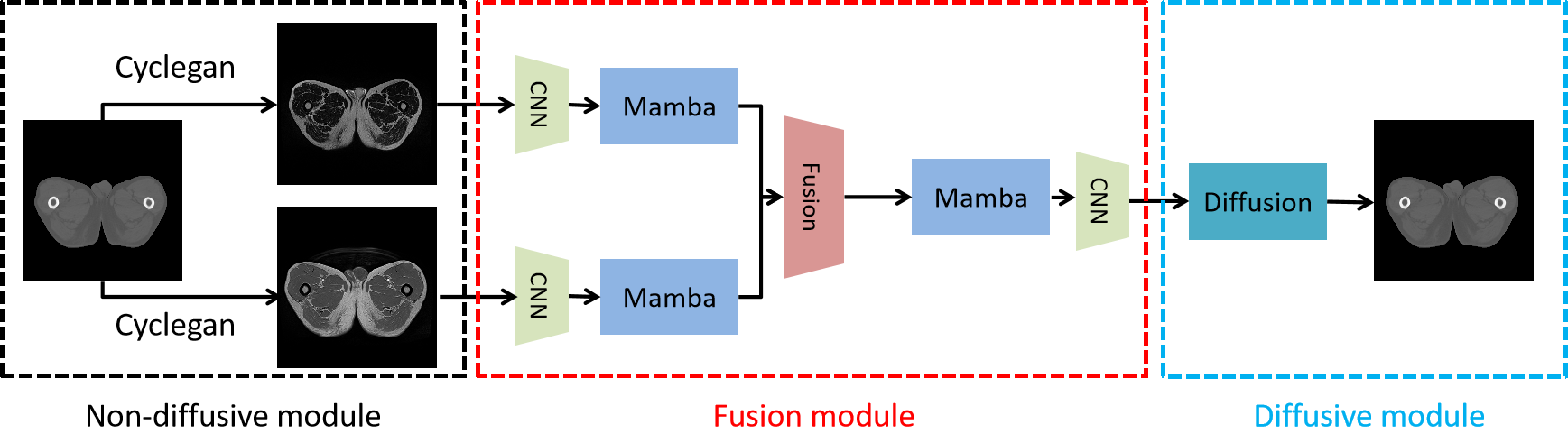}
\caption{The network architecture. It includes three key components: GAN-based generation module, feature fusion module and diffusive module. The process of translating CT into MR is similar.} \label{fig1}
\end{figure*}

\section{Method}
The detailed workflow of our MM2CT model is illustrated in Fig.~\ref{fig1}, showcasing how these components interact to achieve seamless image translation and enhanced feature extraction. 
The translation module is responsible for converting multi-contrast MR images into corresponding CT images, ensuring high-quality and accurate transformations. The multi-modal feature fusion module aims to integrate information from multiple MRI modalities.


\subsection{Translation Module}
The conversion module consists of two key components: a non-diffusive module and a diffusive module. The non-diffusive module initially processes unpaired data from the training set by estimating source images corresponding to target images using ResNet-based generators~\cite{johnson2016perceptual}, thereby establishing a foundation for subsequent transformation. To fully utilize information from both imaging modalities, we introduce a feature fusion module prior to the diffusive module input. The diffusive module based on the UNet backbone then uses the fused multi-modal MRI features as conditional information to guide the generation process of CT images. The diffusive module plays a central role in the conversion process, where the forward diffusion process acts as a low-pass filter, effectively extracting low-frequency information from images, which serves as a crucial starting point for the reverse diffusion process. 

\subsection{Fusion Module}
We introduce a feature fusion module based on the Mamba architecture, which effectively integrates information from T1- and T2-weighted modalities. The fusion module's architecture can be described as follows. Initially, for low-level feature extraction, convolutional layers and Mamba blocks are employed for preliminary processing. Subsequently, we implement shallow channel swapping and deep fusion modules to effectively integrate local detail features from different modalities, ensuring comprehensive extraction of modality specific information. Following these processing steps, the features undergo further processing through Mamba blocks and convolutional layers before generating the synthesized image, which is then forwarded to the subsequent diffusive module.

\begin{figure}[t]
\centering
\includegraphics[width=0.8\linewidth]{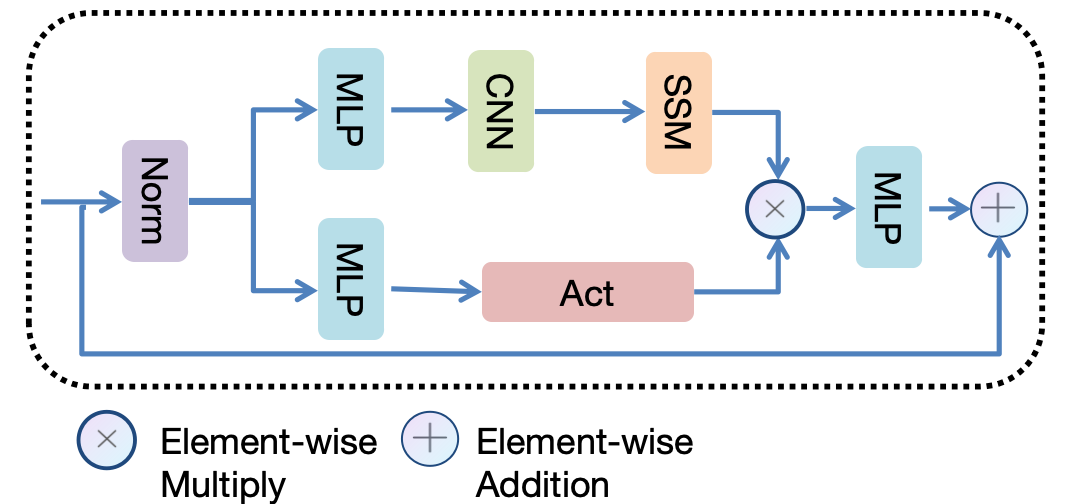}
\caption{Mamba block used in Fusion module} \label{fig2}
\end{figure}

The convolutional layers provide a direct and effective approach for capturing local semantic details. We employ two convolutional layers, each with a 3×3 kernel size and stride of 1. The internal structure of the Mamba block, as illustrated in Fig.~\ref{fig2}, begins with a normalization layer for input processing, followed by a dual-path structure. In the first path, the signal sequentially passes through a linear layer, convolution operation, and SSM module. The second path consists of a linear layer and activation function. The outputs from both paths are merged through multiplication, then processed by a linear layer and added to the original input, forming a residual structure output.

The channel swapping Mamba module represents shallow feature fusion. Inspired by Mambadfuse~\cite{li2024mambadfuse}, we implement alternating channel swapping between the two input feature to achieve lightweight exchange. The channel swapping module facilitates feature interaction between T1 and T2 modalities, establishing inter-modal correlations.

The cross-modal Mamba module is responsible for deep feature fusion. The current Mamba architecture faces challenges when processing multi-modal image information, primarily due to its lack of cross-modal feature fusion capabilities similar to cross-attention mechanisms. To address this limitation, inspired by the concept of cross-attention, we leverage a cross-modal Mamba block~\cite{li2024mambadfuse} designed to facilitate cross-modal feature interaction and fusion. In this approach, we project features from both modalities into a shared space and employ a gating mechanism to encourage complementary feature learning while suppressing redundant features.

\begin{figure}[t]
\centering
\includegraphics[width=\linewidth]{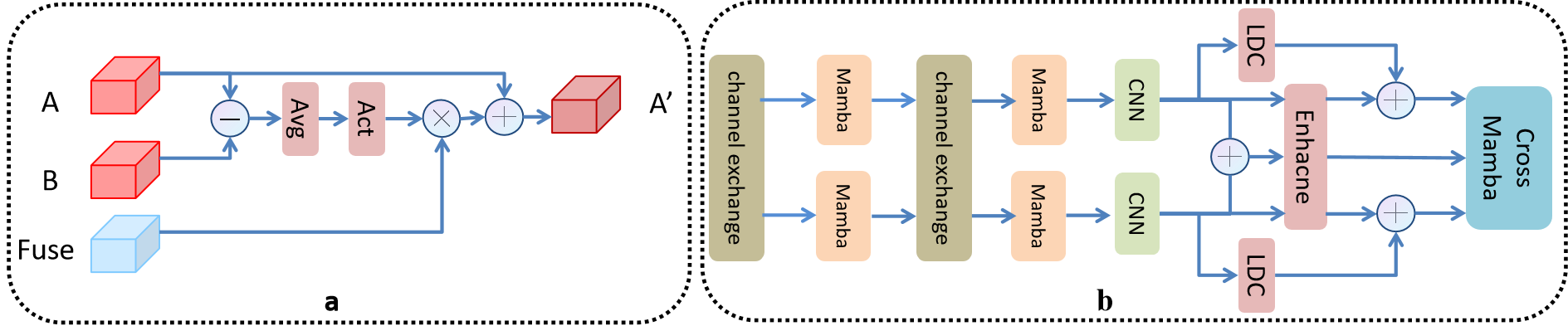}
\caption{Enhance module architecture and Fusion module} \label{fusion}
\end{figure}

Moreover, to better handle inter-modality differences, the fusion feature maps are adjusted based on the disparities between the two modality feature maps to enhance the differences between distinct features. First, a dynamic local convolution mechanism~\cite{huang2022learnable} restores similarities between neighborhoods and enhances texture information perception. Second, a dynamic difference-aware attention mechanism is described as Fig.~\ref{fusion}a. It amplifies subtle differences between input feature maps. The final fusion module is shown in Fig.~\ref{fusion}b.


\subsection{Objective Loss}
In this study, our model is trained by incorporating both traditional adversarial loss and cycle consistency loss, with the discriminator network implemented using a PatchGAN-like architecture ~\cite{isola2017image}. These loss functions jointly constitute the model's overall objective function, ensuring the quality of generated images and maintaining consistent feature mapping between the source and target domains. The overall objective loss function is formulated as follows: 
\begin{equation}
Loss_{all} = \lambda_{l1}Loss_{l1} + \lambda_{gan}Loss_{GAN} + Loss_{Dis}
\end{equation}
where $\lambda_{l1}$ and $\lambda_{gan}$ are the weighting of $Loss_{l1}$ and $Loss_{GAN}$, respectively. 
All parameters used in the loss function are the same as the setting in~\cite{ozbey2023unsupervised}. 

\begin{table}[b]
\centering
\caption{Quantitative Comparison Between On The Pelvis Dataset.}
\label{table}
\begin{tabular}{p{100pt}|p{100pt}|p{30pt}|p{30pt}}
\hline
Modality& Model& PSNR & SSIM\\
\hline
$T1\rightarrow CT$& CycleGAN~\cite{zhu2017unpaired} & 22.69 & 81.47 \\
$T2\rightarrow CT$& CycleGAN~\cite{zhu2017unpaired} & 21.57 & 79.56 \\
$T1\rightarrow CT$& Syndiff~\cite{ozbey2023unsupervised} & 25.36 & 88.97 \\
$T2\rightarrow CT$& Syndiff~\cite{ozbey2023unsupervised}& 23.99 & 87.76 \\
$T1\&T2 \rightarrow CT$& MM2CT & \bfseries25.72 & \bfseries89.54 \\
\hline
\end{tabular}
\label{tab1}
\end{table}

\section{Experiments and Result}
\subsection{Dataset}
We demonstrate our model on a multi-modal pelvic MRI-CT dataset~\cite{nyholm2018mr}. While all unsupervised medical image translation models are trained on unpaired images, performance assessments evaluate on the paired and registered images.  

Pelvic from 15 subjects were analyzed, with a split of (9,2,4) subjects. Elastic registration was performed using the SimpleITK library to register T1 and CT volumes onto T2 volumes in validation and test sets. For T1 scans, TE=7.2ms, TR=500-600ms, 0.10$\times$0.10$\times$3mm$^3$ resolution.
For T2 scans, TE=97ms, TR=6000-6600ms, 0.88$\times$0.88$\times$2.50mm$^3$ resolution.
For CT scans, 0.10$\times$0.10$\times$3mm$^3$ resolution, Kernel=B30f, or 0.10$\times$0.10$\times$2mm$^3$ resolution, Kernel=FC17 were prescribed. We apply the following pre-processing steps: resampling of all volumes and corresponding labels to $1.0\times1.0\times1.0mm^3$.

\subsection{Implement Details}
The networks are trained using an Adam optimizer with a batch size of 4 and 100 epochs. The experiments are implemeted using Pytorch and are trained on an A100 GPU with 80GB of memory. For faithfulness, we employ two widely recognized metrics for quantitative evaluations, namely Peak Signal-to-Noise Ratio (PSNR) and Structural Similarity Index Measure (SSIM). PSNR assesses whether the synthesized images is a uniform projection of the image, and SSIM concentrates on the visible structure of the images.

\begin{figure*}[t]
\centering
\includegraphics[width=1.0\textwidth]{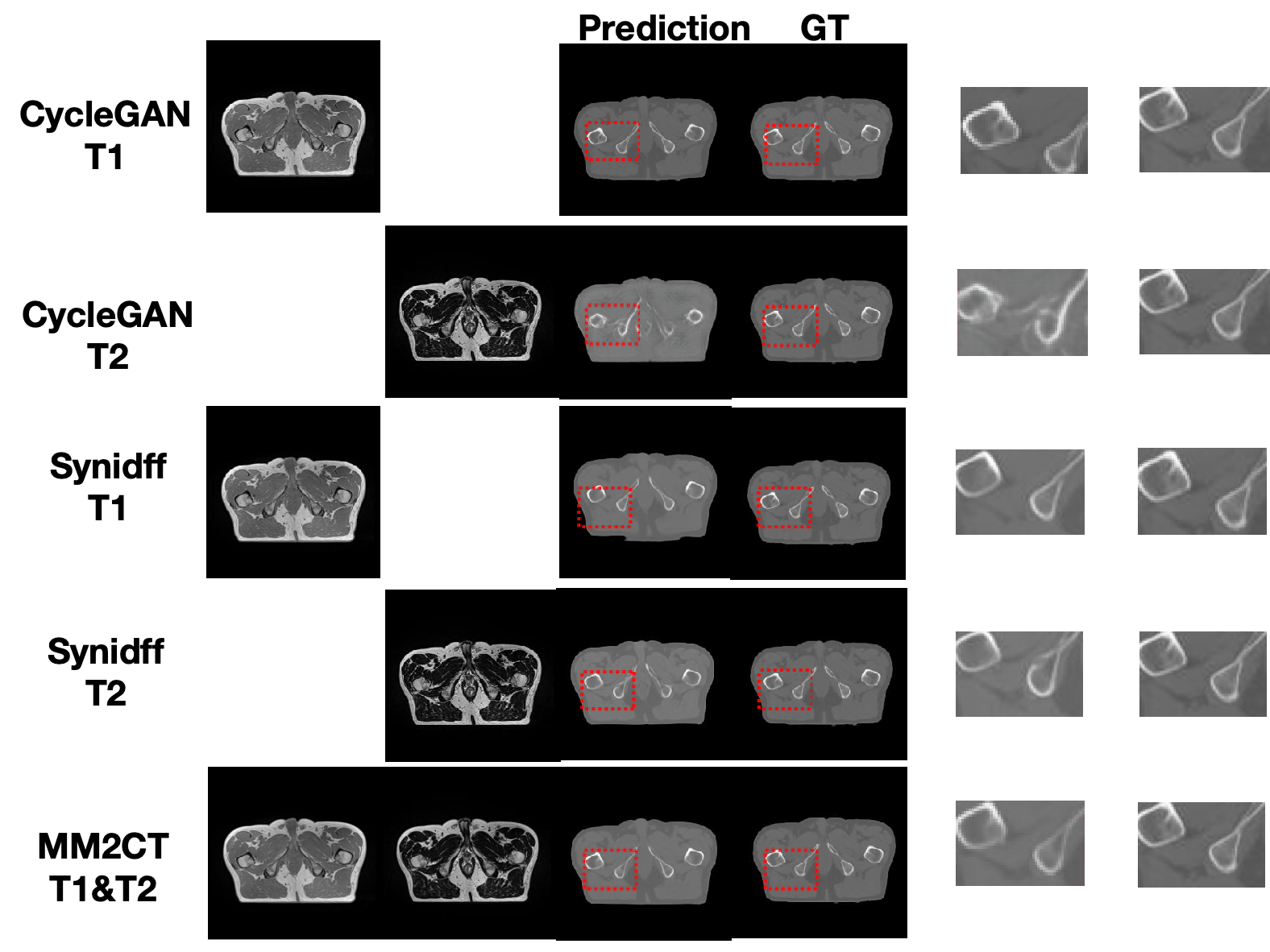}
\caption{The MR-to-CT translation results of different models are shown along with the source image (Source) and the true target (GT). The result demonstrated on the pelvic dataset for mutl-modal MRI-to-CT translation. From left to right, input T1, input T2, synthetic image CT and ground truth image CT.} \label{fig4}
\end{figure*}

\subsection{Results}
\subsubsection{Compare with State-of-the-art Approaches}
The model performance was quantitatively analyzed by comparing the synthesized images with real CT images.
Table~\ref{tab1} presents representative test sample results. Initially, we compared the performance of two single-modality models. The performance of the model is obtained by retraining and testing based on the divided dataset. Syndiff demonstrated the best baseline performance. Further analysis revealed that our proposed model MM2CT significantly outperformed the Syndiff model in both metrics. This superior performance can be attributed to the effective utilization of features from both modalities. The performance of our MM2CT is improved by $0.36$dB PSNR and $0.57\%$ SSIM compared to Syndiff. These results confirm that our model better preserves high-frequency detail information in images, producing synthetic images that more closely resemble the quality of real CT images.
As shown in Fig.~\ref{fig4}, the images generated by MM2CT demonstrate superior detail preservation and more natural texture features. The effectiveness of MM2CT is validated not only through quantitative metrics but also through direct visual quality assessment.

\subsection{Ablation Studies}
To validate the effectiveness of core components in the MM2CT model, we conducted a series of ablation experiments. Our analysis focused on two key modules: the Mamba fusion module and the Dynamic Enhancement (DE) module, examining their impact on the model's overall performance. Since the DE module is integrated within the Fusion block, conducting isolated ablation experiments for this component presented certain challenges.

As demonstrated by the experimental results in Table~\ref{tab2}, the complete MM2CT model significantly outperformed its ablated variants across all evaluation metrics, substantiating the necessity of each component. Specifically, compared to the version without the DE module, the fully-equipped MM2CT achieved improved image quality metrics, with a $0.23$ dB increase in PSNR and $0.09\%$ improvement in SSIM. These experimental findings strongly validate that our proposed complete MM2CT architecture better preserves and reconstructs image details, thereby achieving superior performance in the cross-modal image translation task from MR to CT.

\begin{table}[t]
\centering
\caption{Ablation study on cross Mamba and Dynamic Enhancement (DE) modules.}\label{tab2}
\centering
\renewcommand{\arraystretch}{1}
\setlength{\tabcolsep}{14pt}
\begin{tabular}{c|c|c|c}
\hline
\multicolumn{2}{c}{Configure} &  \multicolumn{2}{|c}{Metrics} \\
\hline
Mamba module& DE & PSNR $\uparrow$ & SSIM $\uparrow$ \\
\hline
- & - &  25.36 & 88.97 \\
$\checkmark$ & - & 25.49 & 89.45\\
$\checkmark$ & $\checkmark$ & 25.72 & 89.54\\
\hline
\end{tabular}
\end{table}

\section{Conclusion}
The proposed MM2CT framework presents an innovative solution for cross-modal synthesis in medical imaging, significantly improving the accuracy of MRI-to-CT image translation. By developing a state-space-based multi-modal feature fusion mechanism, this work establishes novel theoretical foundations for multi-modal image synthesis. Experimental results demonstrate that the framework outperforms current state-of-the-art methods across multiple quantitative metrics, including SSIM and PSNR, validating its potential value and prospects for clinical applications. Future work will focus on extending validation to multi-center datasets to furtheßr enhance generalization and clinical applicability of the proposed model.

\begin{credits}
\subsubsection{\ackname} 
The research in this paper was funded by Inno HK program.
\subsubsection{\discintname}
The authors have no competing interests to declare relevant to this article’s content.
\end{credits}

%
%
%
\bibliographystyle{splncs04}
\bibliography{main.bib}

\begin{thebibliography}{10}
\providecommand{\url}[1]{\texttt{#1}}
\providecommand{\urlprefix}{URL }
\providecommand{\doi}[1]{https://doi.org/#1}

\bibitem{armanious2020medgan}
Armanious, K., Jiang, C., Fischer, M., K{\"u}stner, T., Hepp, T., Nikolaou, K., Gatidis, S., Yang, B.: {MedGAN: Medical image translation using GANs}. Computerized Medical Imaging and Graphics  \textbf{79},  101684 (2020)

\bibitem{atli2024i2i}
Atli, O.F., Kabas, B., Arslan, F., Demirtas, A.C., Yurt, M., Dalmaz, O., {\c{C}}ukur, T.: I2i-mamba: Multi-modal medical image synthesis via selective state space modeling. arXiv preprint arXiv:2405.14022  (2024)

\bibitem{bosch2023risk}
Bosch~de Basea~Gomez, M., Thierry-Chef, I., Harbron, R., Hauptmann, M., Byrnes, G., Bernier, M.O., Le~Cornet, L., Dabin, J., Ferro, G., Istad, T.S., et~al.: Risk of hematological malignancies from ct radiation exposure in children, adolescents and young adults. Nature Medicine  \textbf{29}(12),  3111--3119 (2023)

\bibitem{F2PASeg2025miccai}
Chen, L., Wu, Z., Lei, T., Bai, X., Feng, M., Wang, Y., Meng, G., Lei, Z., Liu, H.: F2paseg: Feature fusion for pituitary anatomy segmentation in endoscopic surgery. In: Medical Image Computing and Computer Assisted Intervention - {MICCAI}, 2025

\bibitem{DBLP:conf/iros/ChenZGL0W0L24}
Chen, Z., Zhang, Z., Guo, W., Luo, X., Bai, L., Wu, J., Ren, H., Liu, H.: Asi-seg: Audio-driven surgical instrument segmentation with surgeon intention understanding. In: {IEEE/RSJ} International Conference on Intelligent Robots and Systems, {IROS} 2024, Abu Dhabi, United Arab Emirates, October 14-18, 2024. pp. 13773--13779. {IEEE} (2024). \doi{10.1109/IROS58592.2024.10801703}, \url{https://doi.org/10.1109/IROS58592.2024.10801703}

\bibitem{dalmaz2022resvit}
Dalmaz, O., Yurt, M., {\c{C}}ukur, T.: Resvit: residual vision transformers for multimodal medical image synthesis. IEEE Transactions on Medical Imaging  \textbf{41}(10),  2598--2614 (2022)

\bibitem{dayarathna2023deep}
Dayarathna, S., Islam, K.T., Uribe, S., Yang, G., Hayat, M., Chen, Z.: Deep learning based synthesis of mri, ct and pet: Review and analysis. Medical Image Analysis p. 103046 (2023)

\bibitem{georgescu2023multimodal}
Georgescu, M.I., Ionescu, R.T., Miron, A.I., Savencu, O., Ristea, N.C., Verga, N., Khan, F.S.: Multimodal multi-head convolutional attention with various kernel sizes for medical image super-resolution. In: Proceedings of the IEEE/CVF winter conference on applications of computer vision. pp. 2195--2205 (2023)

\bibitem{MRtoCT2025isbi}
Gong, C., Huang, Z., Wu, Z., Zhao, M., Xu, F., Bai, X., Feng, M., Granados, A., Meng, G., Lei, Z., Liu, H.: Mr-to-ct translation using frequency-separated diffusion models. In: 2025 IEEE 22nd International Symposium on Biomedical Imaging (ISBI)

\bibitem{huang2022learnable}
Huang, P.K., Ni, H.Y., Ni, Y., Hsu, C.T.: Learnable descriptive convolutional network for face anti-spoofing. In: BMVC. vol.~2, p.~7 (2022)

\bibitem{isola2017image}
Isola, P., Zhu, J.Y., Zhou, T., Efros, A.A.: Image-to-image translation with conditional adversarial networks. In: Proceedings of the IEEE Conference on Computer Vision and Pattern Recognition. pp. 1125--1134 (2017)

\bibitem{johnson2016perceptual}
Johnson, J., Alahi, A., Fei-Fei, L.: Perceptual losses for real-time style transfer and super-resolution. In: European conference on computer vision. pp. 694--711. Springer (2016)

\bibitem{li2022wavtrans}
Li, G., Lyu, J., Wang, C., Dou, Q., Qin, J.: Wavtrans: Synergizing wavelet and cross-attention transformer for multi-contrast mri super-resolution. In: International Conference on Medical Image Computing and Computer-Assisted Intervention. pp. 463--473. Springer (2022)

\bibitem{li2019diamondgan}
Li, H., Paetzold, J.C., Sekuboyina, A., Kofler, F., Zhang, J., Kirschke, J.S., Wiestler, B., Menze, B.: Diamondgan: unified multi-modal generative adversarial networks for mri sequences synthesis. In: Medical Image Computing and Computer Assisted Intervention--MICCAI 2019: 22nd International Conference, Shenzhen, China, October 13--17, 2019, Proceedings, Part IV 22. pp. 795--803. Springer (2019)

\bibitem{li2021medical}
Li, Y., Zhao, J., Lv, Z., Li, J.: Medical image fusion method by deep learning. International Journal of Cognitive Computing in Engineering  \textbf{2},  21--29 (2021)

\bibitem{li2024mambadfuse}
Li, Z., Pan, H., Zhang, K., Wang, Y., Yu, F.: Mambadfuse: A mamba-based dual-phase model for multi-modality image fusion. arXiv preprint arXiv:2404.08406  (2024)

\bibitem{li2022new}
Li, Z., Xia, P., Tao, R., Niu, H., Li, B.: A new perspective on stabilizing gans training: Direct adversarial training. IEEE Transactions on Emerging Topics in Computational Intelligence  \textbf{7}(1),  178--189 (2022)

\bibitem{liu2025vmamba}
Liu, Y., Tian, Y., Zhao, Y., Yu, H., Xie, L., Wang, Y., Ye, Q., Jiao, J., Liu, Y.: Vmamba: Visual state space model. Advances in neural information processing systems  \textbf{37},  103031--103063 (2025)

\bibitem{nyholm2018mr}
Nyholm, T., Svensson, S., Andersson, S., Jonsson, J., Sohlin, M., Gustafsson, C., Kjell{\'e}n, E., S{\"o}derstr{\"o}m, K., Albertsson, P., Blomqvist, L., et~al.: Mr and ct data with multiobserver delineations of organs in the pelvic area—part of the gold atlas project. Medical physics  \textbf{45}(3),  1295--1300 (2018)

\bibitem{ozbey2023unsupervised}
{\"O}zbey, M., Dalmaz, O., Dar, S.U., Bedel, H.A., {\"O}zturk, {\c{S}}., G{\"u}ng{\"o}r, A., {\c{C}}ukur, T.: Unsupervised medical image translation with adversarial diffusion models. IEEE Transactions on Medical Imaging  (2023)

\bibitem{phan2023structure}
Phan, V.M.H., Liao, Z., Verjans, J.W., To, M.S.: Structure-preserving synthesis: Maskgan for unpaired mr-ct translation. In: International Conference on Medical Image Computing and Computer-Assisted Intervention. pp. 56--65. Springer (2023)

\bibitem{qu2024survey}
Qu, H., Ning, L., An, R., Fan, W., Derr, T., Liu, H., Xu, X., Li, Q.: A survey of mamba. arXiv preprint arXiv:2408.01129  (2024)

\bibitem{wu2024towards}
Wu, H., Fu, J., Ye, H., Zhong, Y., Zou, X., Zhou, J., Wang, Y.: Towards multi-modality fusion and prototype-based feature refinement for clinically significant prostate cancer classification in transrectal ultrasound. In: International Conference on Medical Image Computing and Computer-Assisted Intervention. pp. 724--733. Springer (2024)

\bibitem{DBLP:journals/bspc/WuLZWWLLL24}
Wu, Z., Lau, C.Y., Zhou, Q., Wu, J., Wang, Y., Liu, Q., Lei, Z., Liu, H.: Surgivisor: Transformer-based semi-supervised instrument segmentation for endoscopic surgery. Biomed. Signal Process. Control.  \textbf{87}(Part {B}),  105434 (2024). \doi{10.1016/J.BSPC.2023.105434}, \url{https://doi.org/10.1016/j.bspc.2023.105434}

\bibitem{xie2024fusionmamba}
Xie, X., Cui, Y., Tan, T., Zheng, X., Yu, Z.: Fusionmamba: Dynamic feature enhancement for multimodal image fusion with mamba. Visual Intelligence  \textbf{2}(1), ~37 (2024)

\bibitem{FT-Reg2025}
Xu, F., Zhao, M., Wu, Z., Liu, H., Meng, G.: Ft-reg: Unsupervised multimodal medical image registration using dynamic feature translation. Machine Intelligence Research

\bibitem{DBLP:conf/miccai/XuWCCLL24}
Xu, H., Wu, J., Cao, G., Chen, Z., Lei, Z., Liu, H.: Transforming surgical interventions with embodied intelligence for ultrasound robotics. In: Linguraru, M.G., Dou, Q., Feragen, A., Giannarou, S., Glocker, B., Lekadir, K., Schnabel, J.A. (eds.) Medical Image Computing and Computer Assisted Intervention - {MICCAI} 2024 - 27th International Conference, Marrakesh, Morocco, October 6-10, 2024, Proceedings, Part {VI}. Lecture Notes in Computer Science, vol. 15006, pp. 703--713. Springer (2024). \doi{10.1007/978-3-031-72089-5\_66}, \url{https://doi.org/10.1007/978-3-031-72089-5\_66}

\bibitem{zhang2023multi}
Zhang, J., Zhang, S., Shen, X., Lukasiewicz, T., Xu, Z.: Multi-condos: Multimodal contrastive domain sharing generative adversarial networks for self-supervised medical image segmentation. IEEE Transactions on Medical Imaging  \textbf{43}(1),  76--95 (2023)

\bibitem{zhu2017unpaired}
Zhu, J.Y., Park, T., Isola, P., Efros, A.A.: Unpaired image-to-image translation using cycle-consistent adversarial networks. In: Proceedings of the IEEE International Conference on Computer Vision. pp. 2223--2232 (2017)

\end{thebibliography}
%




\end{document}